# Investigation of the atomic and electronic structures of highly ordered two-dimensional germanium on Au(111)

W. Wang and R. I. G. Uhrberg

*Department of Physics, Chemistry, and Biology, Linköping University, S-581 83 Linköping, Sweden*

**Abstract**

Low energy electron diffraction (LEED), scanning tunneling microscopy (STM), and photoelectron spectroscopy have been used to study an ordered structure formed by Ge atoms deposited onto the Au(111) surface. Based on a careful analysis of STM images and LEED patterns, we propose a $\begin{pmatrix} 5 & 0 \\ 8 & -14 \end{pmatrix}$ unit cell for the atomic structure of the Ge layer. Core level data indicate that some Ge atoms diffuse into the Au(111) crystal during annealing after deposition at room temperature. This is further corroborated by angle resolved photoelectron spectroscopy measured for different amounts of Ge remaining after sputtering and annealing. The results of the ARPES study clearly exclude the interpretation, in the literature, of a parabolic band as part of a Dirac cone of germanene.

## I. INTRODUCTION

The unique properties of graphene have attracted many scientists to engage in the field of two-dimensional (2D) materials [1]. In order to find novel graphene-like 2D materials, the elements Si, Ge, and Sn in group IV of the periodic system are considered as promising candidates. Studies of silicene formation have been performed on various substrates, such as Ag(111) [2-4], ZrB$_2$(0001) [5] and Ir(111) [6]. Stanene (Sn) has been investigated on Bi$_2$Te$_3$(111) [7]. In this paper, we focus on germanene (Ge), which is theoretically predicted to have a buckled honeycomb structure in the free-standing form but still possessing Dirac Fermions close to the Fermi level [8]. Germanene is also predicted to have a high charge carrier mobility, spin Hall effect, non-trivial topological properties, etc. [9]. Furthermore, in contrast to graphene, it is relatively easy to open a band gap in the case of germanene, which is of importance in order to realize real applications such as germanene-based field-effect devices [10,11].

Recently, germanene has been claimed to be successfully fabricated on metal surfaces, such as Au(111) [12,13], Al(111) [14,15], Cu(111) [16] and Pt(111) [17]. However, one of the signatures of graphene-like 2D materials, the Dirac cone, has not been presented with convincing experimental evidence yet. Currently, the atomic and electronic structures of Ge/Au(111) are not well known. Au(111) has been suggested as a suitable substrate for the growth of germanene since it is believed that Ge atoms will not form a surface alloy with Au(111). This idea comes from the report that Au atoms do not form an alloy with Ge(111) [18]. However, since Ge-Au alloys do exist, it seems premature to rule out alloy formation for the Ge/Au(111) system. Low energy electron diffraction (LEED) from Ge/Au(111) shows quite complex patterns indicating a mixture of superstructures at monolayer (ML) coverage, which Dávila *et al.* [12] interpreted as a combination of 5×5, √7×√7, and √19×√19 periodicities. With a few Ge layers, LEED was reported to show an 8×8 periodicity while the scanning tunneling microscopy (STM) results showed a different periodicity [13]. Some features of the band structure around the $\bar{\Gamma}$ and $\bar{K}$ points of the 1×1 surface Brillouin zone (SBZ) of Ge/Au(111) were discussed in terms of partially visible Dirac cones in Ref. 13. This issue is addressed in the present paper.

Here, we present LEED, STM and photoelectron spectroscopy (PES) results that lead to a more detailed understanding of the Ge/Au(111) system. Our results indicate that Ge atoms diffuse into the Au(111) crystal during annealing after deposition at room temperature. From our core-level PES studies, after several cycles of sputtering and annealing, we conclude that it is difficult to completely remove Ge atoms from Au(111). LEED patterns of the Ge/Au(111) sample prepared in this study are significantly sharper and show more details than LEED results in the literature.

By comparing with published LEED patterns, we find similarities with the report of "few layer germanene" in Ref. 13. The LEED spots observed in that study constitute a subset of the spots presented here. Our LEED data together with better-resolved STM images included in this paper, lead to a significantly improved picture of the atomic structure of Ge/Au(111). Angle resolved PES (ARPES) data obtained for different amounts of Ge remaining after sputtering and annealing clearly show that a parabolic dispersion observed around the $\bar{\Gamma}$ point is not a Dirac cone of germanene.

## II. EXPERIMENTAL DETAILS

Samples were prepared *in situ* in two separate ultrahigh vacuum (UHV) systems. One was equipped with LEED and STM, and the other was equipped with LEED and PES. A clean Au(111) surface was prepared by repeated cycles of sputtering by Ar$^+$ ions (1 keV) and annealing at approximately 400 °C until a sharp LEED pattern typical of the so called "herringbone" reconstruction (often referred to as a 22×√3 reconstruction) was obtained [19]. About 1 ML of Ge was deposited at a rate of ~0.46 ML/min, while the Au(111) substrate was kept at room temperature. Post annealing at around 300 °C was applied to obtain a sharp LEED pattern and well-defined core level spectra. STM images were recorded at room temperature using an Omicron variable temperature STM in the UHV system at Linköping University. All STM images were measured in constant current mode with a tunneling current of 300 pA. ARPES and core level PES data were obtained at the MAX-lab synchrotron radiation facility using the beam line I4 end station. Data were acquired at room temperature by a Phoibos 100 analyzer from Specs with a two-dimensional detector. The energy and angular resolutions were 50 meV and 0.3°, respectively. Angle integrated Ge 3d and Au 4f core-level spectra were measured at room temperature, using a photon energy of 135 eV.

## III. RESULTS AND DISCUSSION

Only very weak 1×1 LEED spots from the Au(111) substrate were observed after deposition. Subsequent annealing at 300 °C resulted in sharp LEED spots arranged in a complicated pattern, shown in Figs. 1(a) and 1(b). By analyzing the LEED data at different electron energies [20], we propose that the pattern corresponds to a $\begin{pmatrix} 5 & 0 \\ 8 & -14 \end{pmatrix}$ periodicity. The schematic LEED pattern with this periodicity is shown in Fig. 1(c) by white and brown dots. The white dots correspond to the subset of the $\begin{pmatrix} 5 & 0 \\ 8 & -14 \end{pmatrix}$ spots that has been observed in the LEED study at different energies. A similar, but significantly weaker LEED pattern was presented in Ref. 13 for Ge/Au(111). The diffraction spots that can be discerned from the LEED figure in that paper all seem to fit with spots in Figs. 1(a) and 1(b). However, the assignment of the LEED pattern to an 8×8 periodicity find no support from our LEED data.

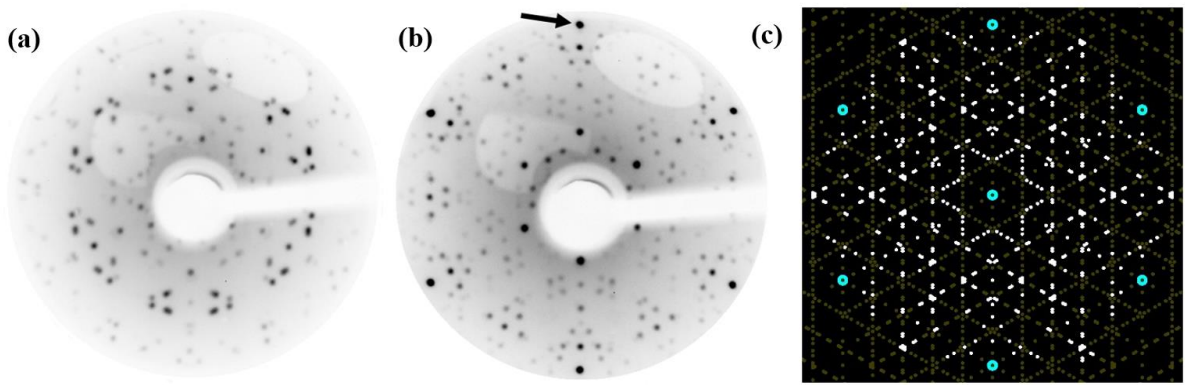

Fig. 1 (color online) (a) and (b) LEED patterns obtained at electron energies of 32 and 53 eV, respectively, from the Au(111) surface after deposition of ~1 ML of Ge. The deposition was done at room temperature at a rate of ~0.46 ML/min and the sample was subjected to post annealing at 300 °C. The LEED data exhibit a complicated pattern of sharp diffraction spots. One of the diffraction spots corresponding to Au(111) 1×1 is indicated by an arrow in (b). (c) Schematic diffraction pattern corresponding to a $\begin{pmatrix} 5 & 0 \\ 8 & -14 \end{pmatrix}$ periodicity. White dots indicate the part of the schematic diffraction pattern that has been identified in LEED patterns at various energies. Diffraction spots corresponding to the brown dots could not be identified in the LEED patterns, which is possibly due to a too low intensity or to vanishing structure factors. Detailed comparisons between experimental and schematic LEED patterns are available in the supplemental materials [20].

Figure 2(a) shows an STM image of the Ge/Au(111) surface exhibiting the LEED pattern shown in Fig. 1. This overview image reveals a striped arrangement of blobs along a $<\bar{1}10>$ direction. From close up images, as that in Fig. 2(b), one finds that the periodicity along the stripes is ~1.41 nm. The separations between the stripes is periodic and can be characterized by alternating long and short distances, as indicated by the white lines, corresponding to an experimentally determined periodicity of ~3.41 nm. A unit cell consistent with the STM images is shown by the blue parallelogram in Fig. 2(b), where the indicated angle is ~93°. These experimental values agree well with those of the $\begin{pmatrix} 5 & 0 \\ 8 & -14 \end{pmatrix}$ unit cell for which the corresponding values are 1.44 nm, 3.50 nm, and 94.7°.

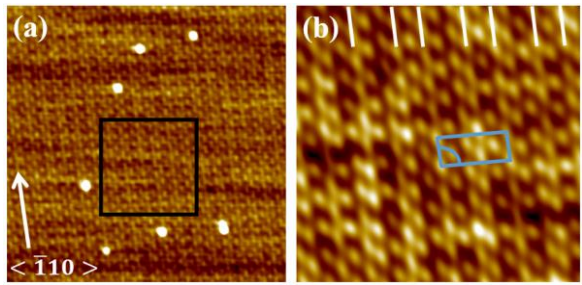

Fig. 2 (color online) (a) Filled state STM image of a 37×37 nm² area of the Ge/Au(111) preparation in Fig. 1. White arrow indicates a $<\bar{1}10>$ direction of the Au crystal. From this overview image, one can perceive a striped ordering along $<\bar{1}10>$. (b) Zoomed-in image of the area inside the black square (~16×16 nm²) in (a). The image has been filtered to remove pixel noise. A unit cell, determined from the periodicity of the STM image, is indicated by a blue parallelogram. This unit cell fits well with a $\begin{pmatrix} 5 & 0 \\ 8 & -14 \end{pmatrix}$ periodicity as verified in the paper. The STM was recorded at room temperature in constant current mode with a tunneling current of 300 pA and a sample bias of -1.30 V.

The STM images in Figs. 2(a) and 2(b) can be perceived as showing hexagons that are stretched along the $<\bar{1}10>$ direction. Distorted hexagons were also reported in Ref. 13 but the deviation from an ideal hexagonal structure was suggested to be an experimental artifact due to lack of thermal drift compensation. In our case, the appearance of "stretched hexagons" are real since the proper thermal drift compensation was applied. Hence, the STM images in Fig. 2 present the actual local density of states distribution on the surface.

Core-level spectroscopy was used to gain further information about the $\begin{pmatrix} 5 & 0 \\ 8 & -14 \end{pmatrix}$ structure of Ge/Au(111). Figure 3 shows spectra of the Au 4f and Ge 3d core levels. After post annealing at ~300 °C for 10 minutes, the Au 4f spectrum, shown in Fig. 3(a) can be fitted by two components labeled, $B_{Au}$ and $I_{Au}$. These components are interpreted as Au bulk and Au-Ge intermixed contributions to the spectrum, respectively. The Ge 3d spectrum after post annealing, shown in Fig. 3(b), is also composed of two components labeled $S_{Ge1}$ and $S_{Ge2}$. After three cycles of sputtering (15 minutes) and annealing (10 minutes, ~400 °C), the Au 4f spectrum (labeled "sputtered" in Fig. 3(a)) shows essentially only a bulk component.

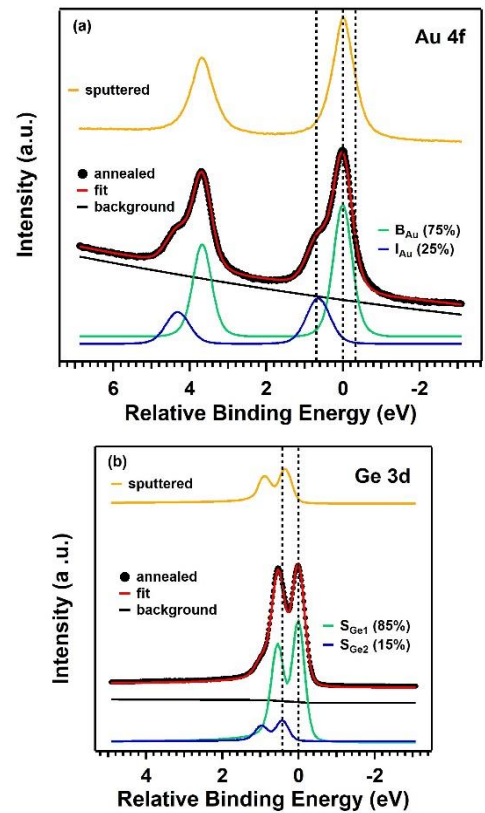

Fig. 3 (color online) (a) Au 4f and (b) Ge 3d core-level spectra obtained at a photon energy of 135 eV at normal emission. The solid circles are the experimental data overlapped by the red fitting curves, which are the sum of two components and a background. (a) $B_{Au}$ and $I_{Au}$ represent Au bulk and Au-Ge intermixed components, respectively. The relative energy of the bulk component is set to 0 eV. Fitting parameters: Spin-orbit split: 3.67 eV, Branching ratio: 0.7, Gaussian widths: 479 and 622 meV, respectively, Lorentzian width: 180 meV. The energy difference between the two components is 0.62 eV. The rightmost dashed line indicates the position of the Au 4f surface component of clean Au(111). (b) $S_{Ge1}$ and $S_{Ge2}$ represent Ge over layer and Au-Ge intermixed components, respectively. Fitting parameters: Spin-orbit split: 0.545 eV, Branching ratios: 0.75 and 0.60, respectively, Gaussian widths: 318 and 372 meV, respectively, Lorentzian width: 110 meV. The asymmetry parameter of the Doniach–Šunjić line profile is 0.05 and the energy difference between the two components is 0.40 eV. The Au 4f and Ge 3d spectra labeled "sputtered" were obtained after three cycles of sputtering and annealing, see the text.

The intensities at the positions of the Au-Ge intermixed component (dashed line at higher binding energy) and of the surface component of clean Au(111) (dashed line at lower binding energy) are negligible [21,22]. The corresponding Ge 3d spectrum showed only one component, which was shifted by approximately -100 meV compared to $S_{Ge2}$. The sputtering and annealing significantly reduced the amount of Ge and the LEED pattern changed drastically. More than 2 hours of additional sputtering and annealing cycles were needed to significantly diminish the Ge 3d signal and to regain emission from the Shockley surface state of Au(111). ARPES results are presented in Fig. 4 for the clean Au(111) sample and for Ge/Au(111) after various steps of sputtering and annealing. Figure 4(a) shows the dispersion of the Shockley surface state around normal emission with a clearly resolved Rashba split. The splitting is ~0.02 Å$^{-1}$ and the bottom of the surface band is located at ~0.5 eV. These values are in nice agreement with both theoretical and experimental results in the literature [23,24].

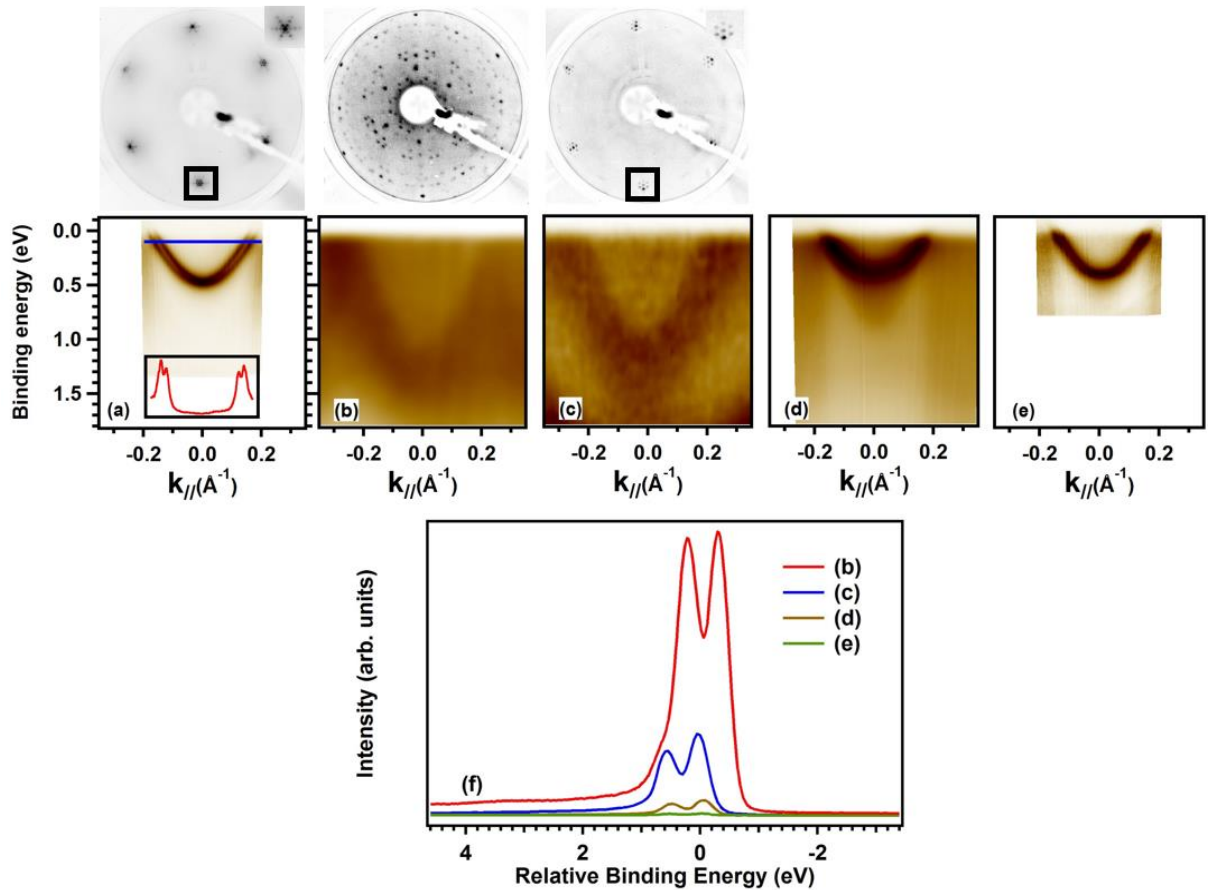

Fig. 4 (color online) (a) – (e) ARPES data along $\bar{K} - \bar{\Gamma} - \bar{K}$ of the 1×1 surface Brillouin zone of Au(111). (a) Shockley surface state of clean Au(111). A Rashba splitting of about 0.02Å$^{-1}$ was determined from the inset image, which is the line profile along the blue line located at 0.1 eV. The upper image shows a typical LEED pattern of clean Au(111). The inset in the upper right coner shows details of the "22×√3" superstructure diffraction around 1×1 spots. (b) ARPES and LEED data after annealing of ~1 ML of Ge deposited onto Au(111). The Shockley surface band is replaced by a parabolic band with a significantly deeper minimum. The LEED pattern of this preparation is consistent with Fig. 1. (c) Parabolic band and LEED pattern after three cycles of sputtering and annealing. The LEED pattern, showing a hexagonal arrangement of satellite spots around 1×1 spots, is completely different from that in (b), while the parabolic band remains essentially unaltered except for a slight upward shift of the band minimum. (d) Parabolic band after seven cycles of sputtering and annealing. The minimum of the parabolic band has moved up close to the minimum of the Shockley surface band, which has now appeared. (e) ARPES data after twelve cycles of sputtering and annealing. The dispersion of the Shockley surface state is essentially recovered but without the clear Rashba splitting displayed in (a) indicating that a small amount of Ge atoms remains. This agrees with the presence of a tiny Ge 3d component, see spectrum (e) in 4(f). (f) Ge 3d spectra, corresponding to the ARPES data in (b)-(e), showing the drop in the Ge 3d emission intensity related to the decrease in the amount of Ge atoms. The photon energies used for PES were 26 eV for (a), 135 eV for (b),(c) and (f) and 27 eV for (d) and (e).

After Ge deposition and post annealing, resulting in the $\begin{pmatrix} 5 & 0 \\ 8 & -14 \end{pmatrix}$ structure, we observed a parabolic band around the $\bar{\Gamma}$ point with a dispersion minimum at ~ 1.4 eV, see Fig. 4(b). This band is of particular interest since a band with similar characteristics was discussed in [13] as a Dirac cone of germanene. Interestingly, the band survived sputtering and annealing as shown in Fig. 4(c). According to the core-level spectra, there was only ~0.15 ML of Ge remaining and the LEED pattern had changed completely. The complicated LEED pattern shown in Fig. 4(b) was replaced by a Au(111) 1×1 pattern with weak satellite spots, forming hexagons with the same orientation as the 1×1 pattern of the substrate. This orientation of the satellite hexagons differs from what is observed for a clean Au(111) surface, see the set of LEED patterns in Fig. 4. The parabolic band gradually disappears, being replace by the Shockley surface band, as the amount of Ge decreases, as illustrated in Figs. 4(b) – 4(e). From these observations, we conclude that the parabolic band cannot possibly be associated with germanene like layers on Au(111). Based on the data presented here, we conclude that the parabolic band originates from intermixing of Ge into the Au(111) substrate. The difficulty to remove the Ge atoms and to recover the Shockley surface band with a well resolved Rashba splitting supports the idea of Ge atom diffusion into the Au crystal.

Our experimental results lead to the following picture of the Ge/Au(111) system as prepared in this study. The initial annealing does not only lead to an ordered structure of the Ge layer but also to intermixing of Ge into the Au crystal. We conclude that the parabolic band observed after annealing does not originate from a germanene like layer on Au(111) since it persists after further sputtering when the LEED pattern has completely changed, with no trace of the ordered $\begin{pmatrix} 5 & 0 \\ 8 & -14 \end{pmatrix}$ superstructure, and the Ge amount is reduced to ~0.15 ML. The difficulty to sputter clean the sample indicates that the remaining Ge atoms reside, to a significant extent, in sub surface layers of the Au substrate.

## IV. SUMMARY

Our combined LEED, STM and PES studies lead to a new understanding of the atomic structure of Ge/Au(111). From these results, it is clear that the atomic structure deviates significantly from the simple 8×8 structure proposed previously [13]. The structure agrees instead with a $\begin{pmatrix} 5 & 0 \\ 8 & -14 \end{pmatrix}$ superstructure with respect to the Au(111) surface. Core-level spectra show clear indications that Ge atoms diffuse into the Au crystal. Furthermore, the interpretation of the parabolic band as a part of a Dirac cone of germanene, can be dismissed since the band persists even when the $\begin{pmatrix} 5 & 0 \\ 8 & -14 \end{pmatrix}$ structure has been removed and the Ge amount is significantly reduced.

## ACKNOWLEDGEMENTS

Technical support from Dr. Johan Adell, Dr. Craig Polley and Dr. T. Balasubramanian at MAX-lab is gratefully acknowledged. Financial support was provided by the Swedish Research Council (Contract No. 621-2014-4764) and by the Linköping Linnaeus Initiative for Novel Functional Materials supported by the Swedish Research Council (Contract No. 2008-6582).

Supplemental material

# Investigation of the atomic and electronic structures of highly ordered two-dimensional germanium on Au(111)

W. Wang[*] and R. I. G. Uhrberg

*Department of Physics, Chemistry, and Biology, Linköping University, S-581 83 Linköping, Sweden*

* Corresponding author: W. Wang, weiwa49@ifm.liu.se

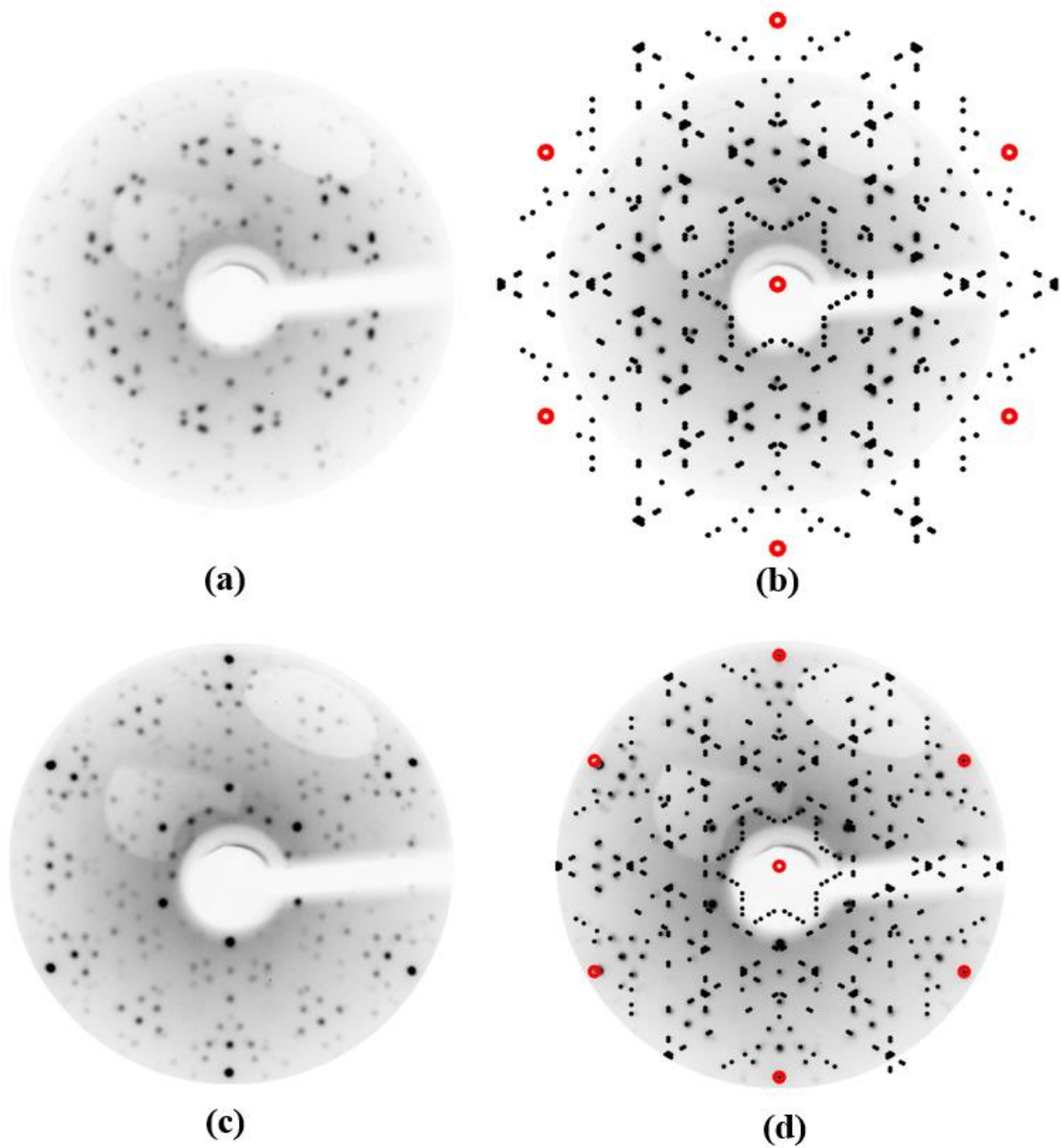

Fig. S1 Experimental and schematic LEED patterns. (a) and (c) LEED patterns obtained at electron energies of 32 and 53 eV, respectively, from the Au(111) surface after deposition of ~1 ML of Ge followed by annealing. (b) and (d) Schematic LEED pattern corresponding to a combination of the six

possible domains of the $\begin{pmatrix} 5 & 0 \\ 8 & -14 \end{pmatrix}$ structure, superimposed on the experimental results in (a) and (c), respectively. Only spots observed experimentally are included in the schematic LEED pattern, c.f., Fig. 1(c). From (b) and (d), one can see that all the discernable experimental diffractions spots are well matched by the dots of the schematic LEED pattern based on a $\begin{pmatrix} 5 & 0 \\ 8 & -14 \end{pmatrix}$ unit cell.